# Neural computation of surface border ownership and relative surface depth from ambiguous contrast inputs

**Birgitta Dresp-Langley**

ICube UMR 7357 CNRS and University of Strasbourg

2, rue Boussingault

67000 Strasbourg

FRANCE

**Stephen Grossberg**

Center for Adaptive Systems

Graduate Program in Cognitive and Neural Systems

Department of Mathematics

Boston University

677 Beacon Street

Boston, MA, 02215

**Abstract**

The segregation of image parts into foreground and background is an important aspect of the neural computation of 3D scene perception. To achieve such segregation, the brain needs information about border ownership; that is, the belongingness of a contour to a specific surface represented in the image. This article presents psychophysical data derived from 3D percepts of figure and ground that were generated by presenting 2D images composed of spatially disjoint shapes that pointed inward or outward relative to the continuous boundaries that they induced along their collinear edges. The shapes in some images had the same contrast (black or white) with respect to the background gray. Other images included opposite contrasts along each induced continuous boundary. Psychophysical results show that figure-ground judgment probabilities in response to these ambiguous displays are determined by the orientation of contrasts only, not by their relative contrasts, despite the fact that many border ownership cells in cortical area V2 respond to a preferred relative contrast. The FACADE and 3D LAMINART models are used to explain these data.

**Introduction**

The non-ambiguous perceptual organization of planar visual images into figure and ground requires the visual system to be able to generate a three-dimensional (3D) representation from a two-dimensional (2D) stimulus input. During viewing of a natural 3D scene, objects that are closer to the viewer may block or occlude the view of objects that are further away. Boundaries of these occluding objects are perceived as belonging to them, a property called *border ownership*. Because occluding objects occur closer in depth than the objects they occlude, border ownership in response to a 3D scene typically coexists with a percept of being closer in depth. The importance of surface border ownership to what may seem nearer to us was already noticed by Galileo (see the review by Dresp-Langley, 2014). The borders of occluding surfaces generally occur in the foreground, while the borders of occluded surfaces generally occur in the background.

      An important problem in visual perception concerns how border ownership assignment occurs in response to 2D pictures, and what role it may play in determining 3D percepts of such pictures. In response to 2D pictures, there are famous examples where the perceptual assignment of surface borders to 3D percepts of foreground and background may be reversible, leading to totally different interpretations of the objects in each representation (Figure 1). Such spontaneous changes in figure-ground perception occur only under particular circumstances due to competition between multiple, approximately balanced, 3D interpretations of the 2D image.

Figure 1

Von der Heydt and his colleagues have published important data from their systematic series of neurophysiological experiments about the border ownership properties of neurons in cortical area V2 of monkeys. In particular, Zhou, Friedman, and von der Heydt (2000) reported data from neurons in cortical area V2 that tend to respond to borders with different firing rates depending on whether the border is owned by an occluding or an occluded surface. These neurons are often maximally excited by a preferred combination of direction-of-contrast and border ownership. Zhang and von der Heydt (2010) further studied the contribution of individual edges to border ownership assignment by decomposing figural contours into fragments. Fragments on the preferred side-of-figure produced facilitation, while fragments on the opposite side produced suppression of neural responses. Border-ownership signals also persist for about a second in the brain (O'Herron and von der Heydt, 2009; 2011). Border-ownership signals are generally consistent over multiple variations in shape geometry, configuration, and contrast (Qiu and von der Heydt, 2005; Qiu, Sugihara, and von der Heydt, 2007; von der Heydt, Qiu, and He, 2003). Fang, Boyaci, and Kersten (2009) furthermore used fMRI and found a border ownership BOLD signal in the human visual cortex.

Grossberg (2015) has proposed a unified explanation of these data properties using the FACADE (Form-And-Color-And-DEpth) model of 3D vision and figure-ground perception, and its further development and extension by the 3D LAMINART laminar cortical model, which together have explained and predicted many data about how the brain consciously sees 3D surface percepts in response to 2D pictures and 3D scenes (Cao and Grossberg, 2005, 2012; Fang and Grossberg, 2009; Grossberg, 1994, 1997, 1999; Grossberg and McLoughlin, 1997; Grossberg and Pinna, 2012; Grossberg, Srinivasan, and Yazdanbakhsh, 2011; Grossberg and Swaminathan, 2004; Grossberg and Yazdanbakhsh, 2005; Grossberg et al., 2008; Kelly and Grossberg, 2000; Leveille, Versace, and Grossberg, 2010; McLoughlin and Grossberg, 1998; Yazdanbakhsh and Grossberg, 2004). As noted above, the von der Heydt et al. data show that various neurons in V2 that are sensitive to border ownership also respond with a preferred contrast polarity. However, the same figure-ground properties can occur in a given configuration when contrast polarities are mixed, or are switched from one polarity to the opposite, across the stimulus fragments that induce 3D surface percepts (e. g. Mathews and Welch, 1997), and the phenomenal "logic" of such shape percepts (see Pinna & Grossberg, 2006) is indeed likely to involve a complex hierarchy of integration levels in the brain.

Early phenomenal descriptions of surface percepts in configurations with illusory contours by Prazdny (1983, 1985) noted that the phenomenal strength of surfaces standing out against uniform backgrounds appears as marked in configurations with inducers of opposite contrast polarites as in configurations with inducers of one and the same polarity. Quantitative data for the relative strength of these percepts were not made available in these earlier reports, however, they were so compelling that they motivated subsequent quantitative accounts for boundary detection mechanisms insensitive to the local sign of contrast elements in the perceptual assignment of border ownership (Grossberg, 1984; Shapley and Gordon, 1985; Grossberg and Mingolla, 1985). Shapley and Gordon studied polarity insensitive surface border detection experimentally in a variety of configurations, which included cases where the resulting surface percept occurred on either side of a perceptual boundary depending on the local direction of contrast, but not on its local sign. Several hierarchical stages of neural integration are at work in the genesis of surface percepts. They involve non-linear integration mechanisms as those suggested by Grossberg and Mingolla (1985) and Shapley and Gordon (1985) well beyond the classic V1 or V2 receptive field and, subsequently, several authors (e.g., Kapadia et al., 1995; Polat & Norcia, 1996; Wehrhahn & Dresp, 1998) reported sign-invariant boundary grouping, sensitive to contrast intensities only, on the basis of local collinear detection facilitation by inducers of opposite polarity (see the recent review by Spillmann, Dresp-Langley and Tseng, 2015).

The postulate that boundary detection by the visual system is insensitive to local variations in contrast polarity was subsequently challenged by findings from studies by He and Ooi (1998), Spehar (2000), and Spehar and Clifford (2003), with new configurations where the contrast polarity varies repeatedly within one and the same inducing element. In these cases, the strength of induced perceptual boundaries (or illusory contours) was found to be significantly diminished, especially at stimulus durations shorter than 300 milliseconds (e.g. Spehar & Clifford, 2003). In contrast with the criteria for variations in contrast sign put forward by Shapley and Gordon (1985), these authors created patterns where the local signs cancel each other out locally, not globally along an axis of boundary induction (see Figure 2 for a schematic overview). These studies hark back to earlier observations on the Ehrenstein illusion (Dresp, Salvano-Pardieu and Bonnet, 1996), where the perceptual strength of the centrally induced surface does not depend on the contrast polarity of the inducing lines provided the contrast sign is homogenous within a given inducing element. When the inducers are fragmented into several parts with variable contrast signs (see Figure 2), we observe considerably weaker surface effects. He and Ooi reported a new ring-shaped illusion, the 'O' illusion (see Figure 2), which is only perceived in fragmented radial lines of one and the same polarity (see again Figure 2). These findings may seem controversial, however, they most of all show that the ways in which contrast polarity variations are locally distributed, and the exposure duration of the stimuli, matter critically in the perceptual genesis of shape illusions. At identical physical luminance, opposite contrast signs within one and the same local inducing element may largely cancel each other out and become less effective in perceptual grouping when viewing durations are not long enough.

Figure 2

Here, we specifically tested for figure-ground assignment in terms of what is seen as standing out "in front" and what is seen as as "lying behind" by creating configurations that in every respect match the criteria of Shapley and Gordon (1985) for sign-invariant boundary induction: inducers of varying sign were displayed on either of two sides of a perceptual boundary while the contrast sign within one and the same inducing element was always homogenous (Figure 3). In these configurations, the orientation, direction and polarity of contrast are locally controlled, and may be mixed or switched from one direction and/or polarity to the opposite (e.g. Shapley & Gordon, 1985; Dresp, 1997) across the stimulus elements that produce the resulting figure-ground percept. The duration of presentation was not limited in time, as in natural free viewng conditions. A key variable of the FACADE theory relative to the orientation of surface-inducing contrast edges was tested by presenting inducing elements with outward-oriented contrast edges (upper panel of configurations in Figure 3) and inducers with inward-oriented edges. The FACADE and 3D LAMINART theories make the clear prediction that only the orientation of the local contrasts, not their sign, determines the surface border assignment and thereby the direction of the resulting figure-ground segregation. We employed an alternative forced choice task similar to that from earlier studies (Dresp, Durand & Grossberg, 2002; Dresp-Langley & Reeves, 2012, 2014).

**Materials and method**

The psychophysical experiments were conducted in accordance with the Declaration of Helsinki (1964) and with the full approval of the corresponding author's institutional (CNRS) ethics committee. Informed written consent was obtained from each of the participants of the psychophysical experiments. Experimental sessions were run under laboratory conditions of randomized free trial-by-trial image viewing using a Dell PC computer equipped with a mouse device and a high resolution color monitor (EIZO LCD 'Color Edge CG275W'). This screen has an in-built calibration device which uses the Color Navigator 5.4.5 interface for Windows. The images were generated in Photoshop using selective combinations of Adobe RGB increments to generate contrast inputs (see also Dresp-Langley, 2015). The luminance levels for each RGB triple could be retrieved from a look-up table after calibration and the values were also cross-checked on the basis of standard photometry using an external photometer and adequate interface software (Cambridge Research Instruments).

*Subjects*

Ten unpracticed observers, mostly students in computational engineering who were unaware of the hypotheses of the study, participated in the experiments. All subjects had normal or corrected-to-normal visual acuity.

*Stimuli*

The stimuli (Figure 3) consisted of six images with different edge contrast inputs. The luminance of the background was 50.5 cd/m$^2$ (148,148,148 RGB) in all eight images. The luminance of the black contrast fragments was 1.5 cd/m$^2$ (0,0,0 RGB) and the luminance of the white contrast fragments was 99.5 cd/m$^2$ (255,255,255 RGB), yielding perfectly balanced Weber contrasts ($L_{feature}-L_{background}/L_{background}$) of -0.97 and 0.97 for negative and positive polarities in the six images with the fragmented edge contrasts. The height of the central surfaces was 10 cm on the screen, whereas the width was 12 cm. In the six images with the ambiguous fragmented edge contours, about 50% of the inner surface contour was void of a contrast, so that 50% of the boundary contour had to be completed perceptually (Dresp, 1997).

*Task instructions*

A classic psychophysical forced choice procedure with three response alternatives was used to measure perceptual decisions for relative depth (figure-ground). Observers were asked to indicate whether the central surface appeared to "stand in front" of', to "lie behind", or to be in the "same plane" as the surrounding surface. It was made sure that all observers understood the instructions correctly before an experimental session was initiated.

*Procedure*

Subjects were seated at a distance of 1 meter from the screen and asked to look at the center of the screen. The experiments were run in a dimmed room (mesopic conditions), with blinds closed on all windows. The six images were presented in random order for about one second each, and each image was presented four times in a session. Inter-stimulation intervals were measured. They typically varied from one to three seconds, depending on the observer, who initiated the next image presentation by striking a key on the computer keyboard. The experiment produced a total of 300 observations from 30 trials per subject in an individual session.

**Results**

The individual data from this depth judgment experiment were analyzed in terms of conditional response frequencies, or the frequencies with which the different perceptual responses ("in front", "behind", "same plane") occurred within a given experimental condition. These frequency distributions, permit conclusions relative to event saliency and allow plotting probabilities (e.g., Overall & Brown, 1957), based on the assumption that a similar frequency distribution is statistically likely to occur in any study population with the same characteristics as the sample population selected for this experiment. To assess whether the observed differences between the response frequencies reflecting the most salient events were statistically predictable, we fed the frequency distributions for "in front" and "behind", which reflect complementary dimensions of the underlying psychological decision, into analysis of variance (ANOVA) using *Systat 11* (see also Dresp, Durand & Grossberg, 2002, or Dresp-Langley & Reeves, 2012 and 2014). The balanced 2x3 factorial design, with stimuli presented in random order, allowed for generation of psychophysical judgements from an even number of independent forced-choice trials per factor level. Criteria for parametric testing, including normality and egality of variance of the frequency distributions, were met.

*Experimental results*

The results (Figure 4) show that the configurations generate a higher event probability for the central surface to be perceived as figure ("in front") when the local contrast edges of the fragmented contour elements are inward directed, as indicated by the distribution of the response frequencies *RF,* with the following average values: $RF_{(in\ front)}$ = 0.83 (SEM = 0,05), $RF_{(behind)}$ = 0.07 (SEM = 0,03) and $RF_{(same)}$ = 0.10 (SEM = 0,04). The configurations generate a higher event probability for the central surface to be perceived as ground ("behind") when the local edges are outward directed: ($RF_{(in\ front)}$ = 0.06 (SEM = 0,02), $RF_{(behind)}$ = 0.75 (SEM = 0,03), $RF_{(same)}$ = 0.19 (SEM = 0,04). These perceptual decisions do not depend on the contrast signs of the local edges. Configurations with negative like-contrasts, positive like-contrasts and mixed contrast polarities produced similar response frequency distributions, with average values as follows: $RF_{(in\ front)}$ = 0.51 (SEM = 0,14), $RF_{(behind)}$ = 0.48 (SEM = 0,12) and $RF_{(same)}$ = 0.10 (SEM = 0,04) for negative like-contrasts; $RF_{(in\ front)}$ = 0.42 (SEM = 0,14), $RF_{(behind)}$ = 0.43 (SEM=0,13) and $RF_{(same)}$ = 0.15 (SEM=0,04) for positive like-contrasts; $RF_{(in\ front)}$ = 0.43 (SEM = 0,11), $RF_{(behind)}$ = 0.42 (SEM = 0,10), $RF_{(same)}$ = 0.15 (SEM = 0,05) for mixed polarities.

ANOVA on the response frequencies for "in front" and "behind" for the two levels of the factor "contrast edge direction" and the three levels of the factor "contrast sign" returned statistically significant effects of "contrast edge direction" on perceptual decisions for "in front" ($F(1,2)$ = 228.30, $p<.001$) and "behind" ($F(1,2)$ = 212,77, $p<.001$). As expected (e.g. Dresp, Durand and Grossberg, 2002), no effect of contrast sign on either type of perceptual decision ($F(1,2)$=2.58, NS on response frequencies for "in front" and $F(1,2)$=0.25, NS on response frequencies for "behind") was observed.

Figure 3

**Discussion**

A unified mechanistic explanation can be given of these various percepts using FACADE and 3D LAMINART model mechanisms (Figures 4 and 5). The motivation of the present experiment was to put two major functional assumptions of the FACADE and LAMINART models to the test: 1) the major determining influence of direction of filling-in on the process that leads to figure-ground and 2) the insensitivity of this process to the polarity of contrast of the inducing elements of the visual configuration. Both assumptions are verified by the experimental data, which show effects consistent with both predictions. The original model mechanisms are summarized here with enough detail to achieve a self-contained exposition.

Figure 4

*Bipole boundary completion can pool over opposite contrast polarities*

In response to all of the images from our experiments here, boundaries can be completed inwardly between pairs of adjacent colinear inducers. The completion process uses the oriented long-range horizontal cooperation of bipole grouping cells in layer 2/3 of cortical area V2, balanced by shorter-range disynaptic inhibition (Figures 5 and 6a). Bipole cells can complete boundaries in response to colinear inducers with the same relative contrasts with respect to the background as well as between inducers with opposite relative contrasts with respect to the background, as shown repeatedly in psychophysical experiments (e.g. Wehrhahn & Dresp, 1998; Tzvetanov & Dresp, 2002). This is true because bipole cells receive their inputs, after several stages of additional processing, from complex cells in layer 2/3 of cortical area V1 (Figures 4 and 5). Complex cells, in turn, pool inputs from simple cells in layer 4 of V1 that have the same preferences for position and orientation, but opposite contrast polarities. As a result, bipole cells can complete boundaries around objects that lie in front of textured backgrounds whose relative contrasts reverse along the perimeter of the object. In the present cases, bipole cells complete rectangular boundaries that abut all their inducers.

Figure 5

*Bipoles are sensitive to T-junctions*

The long-range cooperation and short-range competition processes whereby bipoles complete boundaries are sensitive to any T-junctions that lie along the boundaries that they complete (Figure 6a). In the images with incomplete boundaries, there are no explicit T-junctions in the image. However, when a rectangular boundary is completed, T-junctions are created at the corners of the colinear inducing contrasts. The bipole cells that lie along the orientation of a completed boundary (the "head" of the T) get more excitatory input than do the bipole cells that lie near the head of the T, but whose orientational preference is along the perpendicular or oblique orientation of the inducing contrast (the "stem" of the T). This is true because the bipole cells that are activated along the head of the T receive strong excitatory inputs from both sides of their receptive fields, whereas the bipole cells that are activated along the stem of the T receive strong excitatory inputs from just one side of their receptive fields (Figure 6a). The more strongly activated bipole cells inhibit surrounding bipole cells more than conversely through a spatially short-range competitive network. As a result, the bipole cells near the head that are along the stem get inhibited. An *end gap* hereby forms in each boundary near where the stem of a T touches its head (Figure 6a).

Figure 6

Because the bipole cells can complete rectangular boundaries in response to spatially disjoint inducers with the same relative contrasts with respect to their surrounding regions, or in response to combinations of inducers with opposite relative contrasts, end gaps at the T-junctions can form in either case.

As originally explained in Grossberg (1994, 1997), and simulated in such articles as Kelly and Grossberg (2000), Grossberg and Swaminathan (2004), and Grossberg and Yazdanbakhsh (2005), end gaps trigger a process of figure-ground perception and border ownership in which the rectangular boundaries are perceived in front of the regions that they enclose, which are themselves perceived as a ground at a slightly further depth. For example, the percepts of the Necker cube (Figure 7b; Grossberg and Swaminathan, 2004) can be explained in this way, as can the way that shifts in attention can make an attended disk in Figure 7c look both nearer and darker (Grossberg and Yazdanbakhsh, 2005; Tse, 2005). These concepts are reviewed and extended below in order to explain the conscious 3D surface percepts that are generated by the images from our experiment here.

In order to motivate these theoretical explanations, it is useful to ask the following question: If it is indeed the case that these figure-ground relationships do not depend on having inducers with the same contrast polarity, then why do so many cortical area V2 cells that are sensitive to border ownership also exhibit a particular contrast preference; e.g., Zhou, Friedman, and von der Heydt (2000). This can be understood by going into more detail about how end gaps trigger figure-ground perception and border ownership.

Figure 7

*Feedback between boundaries and surfaces achieves complementary consistency*

The FACADE and 3D LAMINART models (Figures 4 and 5) detail how the figure-ground perception process utilizes feedback between the boundary completion process in the interblob cortical stream and the surface filling-in process in the blob cortical stream within V1, V2, and V4 of visual cortex, This feedback enables boundaries and surfaces to generate a consistent percept, despite the fact that they obey computationally complementary laws. This property is called *complementary consistency*. As will be noted shortly, the mechanisms that ensure complementary consistency also contribute to 3D figure-ground separation. Grossberg (2015) explains in detail how the data of von der Heydt et al. about border ownership and related properties of V2 cells fit into this larger explanatory theory.

In particular, the completed boundaries with their end gaps are projected topographically from the interstripes, or pale stripes, of V2, at which boundaries are completed, to the thin stripes of V2, at which one stage of surface filling-in occurs. When surface filling-in occurs within these boundary inducers, brightness and color can flow out of the end gaps, thereby equalizing the filled-in brightness and color on both sides of the remaining boundaries near these gaps (Figure 7). Only the boundary of the rectangle is closed, so only it can fully contain its surface-filling in. However, in these images, the regions both inside and outside the rectangles are surrounded by closed boundaries, since the frame of the image provides another closed boundary that can contain filling-in between it and the bipole-generated rectangular boundary that lies within it. The significance of this fact will be discussed below.

Figure 8

*Closed boundaries, surface contours, and boundary pruning*

As filling-in occurs, feedback can occur from the surfaces in the thin stripes to the boundaries in the interstipes (Figure 8). These feedback signals occur from each active Filling-In DOmain, or FIDO. They are *surface contours* that are generated by contrast-sensitive on-center off-surround networks that act across position and within the depth represented by each FIDO. These contrast-sensitive networks sense sufficiently large and steep spatial discontinuities in the filled-in brightnesses or colors within their FIDO. They hereby generate surface contour output signals only at the surface positions that are surrounded by closed boundaries. In response to the incomplete inducers in the top row of our experimental stimuli, these regions lie on both sides of the completed boundaries. However, due to the end gaps, surface contour signals are not generated at the boundary positions of the inducers themselves.

The surface contour output signals hereby generate topographic feedback signals to a subset of the boundary representations that induced them (Figure 8). These feedback signals are delivered to the boundary representations via an on-center off-surround network whose inhibitory off-surround signals act within position and across depth (Figure 8). The on-center signals strengthen the boundaries that generated the successfully filled-in surfaces at the same depth, whereas the off-surround signals inhibit spurious boundaries at the same positions but further depths. This inhibitory process is called *boundary pruning*. Surface contour signals hereby strengthen consistent boundaries and prune, or inhibit redundant boundaries.

Because surface signals are generated by the contrasts of a filled-in surface, see for example the two-polarity inducing surfaces used by Spehar and Clifford shown in Figure 2, the surface signals are sensitive to a particular contrast, not to the opposite one. Their feedback to boundaries thus makes the responses of the recipient bipole cells also sensitive to this contrast, even though the bipole cells, in the absence of surface contour feedback signals, respond to both contrast polarities, due to their inputs from V1 complex cells, so that they can complete boundaries of objects in front of textured backgrounds. Thus, after surface contour signals act, their target bipole cells also exhibit sensitivity to a particular contrast polarity, as in the neural data of Zhou, Friendman, and von der Heydt (2000).

In response to 3D scenes, boundary pruning is part of the process of *surface capture* whereby feature contours can selectively fill-in visible surface qualia at depths where binocular fusion of object boundaries can successfully occur, thereby contributing to the formation of closed boundaries that can contain the filling-in process. Surface contour and boundary pruning signals hereby work together to generate 3D percepts based on successfully filled-in surface regions.

For example, the open boundary at Depth 2 in V1 and the V2 pale stripes of Figure 8 can be created due to a monocularly viewed vertical boundary that is seen by only one eye, as occurs during daVinci stereopsis (Cao and Grossberg, 2005; Gillam, Blackburn, and Nakayama, 1999; Nakayama and Shimojo, 2000), and by a pair of horizontal boundaries that do not give rise to strong binocular disparities. Such depth-nonselective boundaries are projected to all depth planes along the line of sight (Cao and Grossberg, 2005; Grossberg and Howe, 2003). The closed boundary at Depth 1 in Figure 8 is due to these boundaries plus a left vertical boundary that is formed at that depth due to binocular disparity matching between the two eyes. As a result of surface filling-in within V2 thin stripes and the formation of surface contours only at Depth 1, the closed boundary at Depth 1 is strengthened, whereas the spurious open boundary at Depth 2 is inhibited.

Figure 9

*From boundary pruning to figure-ground separation*

Remarkably, by eliminating spurious boundaries, the off-surround signals that are activated by surface contours also enable figure-ground separation to proceed. They do so by separating occluding and partially occluded surfaces onto different depth planes, after which partially occluded boundaries and surfaces can be amodally completed behind their occluders. For example, the three rectangles in Figure 9a are perceived as a vertical rectangle in front of a partially occluded horizontal rectangle. Due to the action of surface contours, the redundant copy of the vertical rectangle at a further depth (denoted by D2 in Figure 9a) is inhibited, thereby enabling the horizontal boundaries corresponding to the smaller rectangles to be colinearly completed within depth D2. In response to the picture in Figure 9b, the redundant vertical rectangular boundary is inhibited at depth D2, thereby restoring the boundary fragments at depth D2 that previously were inhibited by the D2 vertical boundaries at end gaps. For this reason, end gaps are not seen in the final depthful percept.

*How the disparity filter eliminates some spurious boundaries in the near depth*

Although the boundaries containing end-gaps (Figure 9a) are eliminated by surface contours at the further depth D2, they are not eliminated in this way from depth D1. These near depth boundary fragments are eliminated by the disparity filter (Figure 7), an inhibitory circuit in V2 that operates along the line of sight and across depth to help solve the correspondence problem (Cao and Grossberg, 2005; Grossberg and Howe, 2003; Grossberg and McLoughlin, 1997). The D1 near depth end gap boundary is inhibited by the D2 far depth rectangular boundary at corresponding positions by the disparity filter, because the latter boundary, being closed, is strengthened by surface contour signals, whereas the former boundary is not. Hence the D2 boundary can inhibit the D1 boundary more than conversely.

Although the disparity filter can eliminate the near depth end gap boundary in response to the image in Figure 9a, it cannot do so in response to the image in Figure 9b. This is because the D2 far depth boundary is not closed in this case, and thus is not strengthened by surface contour feedback signals. The same kind of situation occurs in response to the fragmented inducers from our experiment here. How, then, are end gap near-depth boundaries eliminated in this case.

*From unoccluded and occluded recognition in V2 to unoccluded seeing in V4*

In order to explain how these spurious boundaries are also eliminated, it needs to be explained how additional mechanisms generate the modal, or consciously visible, percepts of the unoccluded parts of both occluding and occluded objects in depth. FACADE theory proposes how boundaries and surfaces may be amodally completed in V2 for purposes of recognition, but also that conscious qualia of the unoccluded surfaces of opaque objects are predicted to be represented in V4. These proposed V2 and V4 representations enable the brain to complete the representations of partially occluded objects behind their occluders for purposes of object recognition, without forcing all occluders to appear transparent, which would be the case if the completed boundaries and surfaces that are illustrated in Figure 9a could generate visible surface qualia. How these V2 and V4 mechanisms may cooperate to achieve both effective recognition and seeing were first described in Grossberg (1994, 1997) and then further developed and simulated in many further articles; e.g., Fang and Grossberg (2009) and Kelly and Grossberg (2000). Grossberg and Yazdanbakhsh (2005) additionally explained and simulated how both opaque and transparent percepts can be generated using the same model mechanisms.

Before summarizing these V2-to-V4 mechanisms for conscious seeing, it is worth noting here that surface contour signals also help to control where the eyes look and to thereby help to regulate how the brain learns invariant object categories. The first role arises because surface contour signals are strongest at the distinctive features of an attended object, such as at high curvature positions along a boundary. In addition to the (thin stripe)-to-(pale stripe) feedback that enhances some boundaries while pruning others, a parallel pathway, that is predicted to occur through cortical area V3A, clarifies how these enhanced surface contour positions can also determine target positions of eye movements that explore an attended object's surface. In all, these signals are proposed to determine where the eyes will look next on an attended surface, and thereby enable inferotemporal cortex to learn view-, size-, and positionally-invariant object categories as the eye movements explore this surface. Thus, the 3D LAMINART model is part of a more comprehensive 3D ARTSCAN Search architecture for active vision wherein 3D boundary and surface representations help to control eye movements for attending, seeing, searching, learning, and recognizing invariant object categories (Cao, Grossberg, and Markowitz, 2011; Chang, Grossberg, and Cao, 2014; Fazl, Grossberg, and Mingolla, 2009; Foley, Grossberg, and Mingolla, 2012; Grossberg, 2009; Grossberg, Srinivasan, and Yazdanbaksh, 2014).

*Boundary enrichment and surface pruning in V4*

To set the stage for explaining these V2-to-V4 processes, keep in mind that the boundary pruning process spares the closest surface representation that successfully fills-in at a given set of positions, while removing redundant copies of the boundaries of occluding objects that would otherwise form at further depths. This process illustrates "the asymmetry between near and far". When the competition from redundant occluding boundaries is removed, the boundaries of partially occluded objects can be amodally completed behind them on boundary copies that represent further depths. Moreover, when the redundant occluding boundaries collapse, the redundant surfaces that they momentarily supported collapse as well. Occluding surfaces are hereby seen to lie in front of occluded surfaces.

These surface representations in V2 are depth-selective due to their depth-selective capture by binocular boundaries, but they do not combine brightness and color signals from both eyes (Figure 4). They are said to be computed within *monocular* Filling-In-DOmains, or FIDOs. The computation of binocular surfaces that combine brightness and color signals from both eyes takes place in V4 (Figure 4). These networks are called *binocular* FIDOs. Here monocular surface signals from both eyes are binocularly matched (pathways 8). The successfully matched binocular signals are pruned by inhibitory signals from the monocular FIDOs. These su*rface pruning* inhibitory signals eliminate redundant feature contour signals at at their own positions and further depths. As a result, occluding objects cannot redundantly fill-in surface representations at multiple depths. This surface pruning process is a second example of the "the asymmetry between near and far".

As in the case of the monocular FIDOs, the feature contour signals to the binocular FIDOs can initiate filling-in only where they are spatially coincident and orientationally aligned with binocular boundaries. Boundary pathways 10 in Figure 5 hereby carry out depth-selective surface capture of the binocularly matched feature contour signals that survive surface pruning. In all, the binocular FIDOs fill-in feature contour signals that: (a) survive within-depth binocular feature contour matching (via pathways 8) and across-depth feature contour inhibition (via pathways 9); (b) are spatially coincident and orientationally aligned with the binocular boundaries (pathways 10); and (c) are surrounded by a connected boundary, or fine web of such boundaries.

Figure 10

In addition, at the binocular FIDOs, the binocular boundaries of nearer depths are added topographically to those that represent further depths (e.g., Figure 10b). This third instance of the asymmetry between near and far is called *boundary enrichment*. These enriched boundaries prevent opaque occluding objects, such as the vertical rectangle in Figure 10c, from looking transparent by blocking filling-in of occluded objects behind them, such as the horizontal rectangle in Figure 10c.

The total filled-in surface representation across all binocular FIDOs—after all three processes of boundary pruning, surface pruning, and boundary enrichment act—represents the visible surface percept. It is called a FACADE representation because it combines properties of Form-And-Color-And-DEpth. As to the three asymmetries between near and far, it is possible that they arise during development due to the asymmetric optic flows that are caused by moving forwards much more than backwards.

*Top-down attention from V4 to V2 eliminates end gap boundaries*
Contour-sensitive top-down feedback from the V4 filled-in surfaces to their generative V2 boundaries obeys the ART Matching Rule (e.g., Carpenter and Grossberg, 1987, 1991), which predicts how top-down object attention works. The ART Matching Rule is defined by a modulatory on-center, off-surround network supported by psychological and neurobiological evidence. There is a convergence about the mathematical form that the rule should take (see Grossberg, 2013, for a review). In the present instance, the modulatory on-centers at each depth, D1 and D2, can strengthen the boundaries that generated the corresponding filled-in surface, while inhibiting other boundaries in its broad off-surround. One consequence of this inhibition is elimination of the spurious end gap boundary at depth D1 (Figure 10d).

The 3D boundary and surface representations that are depicted in Figures 9 and 10 provide an explanation of how the fragmented images from our experiment, each of which is caricatured by the image in Figure 9b, generate their depthful figure-ground percepts, notably why the relative depths of figure and ground depend on the positions of the T-junctions relative to the completed boundaries, but not on the relative inducer contrasts that caused them. In response to the fragmented images, these boundaries need to be completed by bipole grouping cells before T-junctions can be created at the fragmented inducers. Once that happens, surface-filling in within closed boundaries ensues. Figures 9 and 10 clarify how the boundary and surface representations within V2 can lead to recognition of figure and ground objects in V2, without these representations also leading to visible surface qualia. The filled-in surface representations within V4 are predicted to support conscious percepts of the qualia of the unoccluded parts of opaque surfaces. Both unique and bistable transparent percepts can also be explained by these FACADE and 3D LAMINART mechanisms, as has been shown by Grossberg and Yazdanbakhsh (2005).

**Conclusions**

This article presents additional experimental evidence to complement the fact that many cells in cortical area V2 that are sensitive to border ownership, and thus implicated in the process of figure-ground perception, also exhibit a preferred contrast polarity. The experimental results here with configurations that match previously established criteria for sign-invariant boundary grouping show that contrast polarity is often unimportant in determining what part of a 2D picture generates a 3D percept of a closer figure, and what part generates a 3D percept of a further background. Both same-polarity and mixed-polarity sets of figural inducers, with either darker or lighter contrasts compared to the background, can generate the same percepts of relative depth. The results support the hypothesis that V2 is just one stage in a cortical hierarchy that also includes V4 in the generation of surface percepts with figure-ground properties. Mechanisms from FACADE theory and the 3D LAMINART model explain the experimental data here, and all the key V2 data that have been reported in an important series of groundbreaking neurophysiological experiments from the von der Heydt laboratory.

# References

Cao, Y., and Grossberg, S. (2005). A laminar cortical model of stereopsis and 3D surface perception: Closure and da Vinci stereopsis. *Spatial Vision*, 18, 515-578.

Cao, Y., and Grossberg, S. (2012). Stereopsis and 3D surface perception by spiking neurons in laminar cortical circuits: A method of converting neural rate models into spiking models. *Neural Networks*, 26, 75-98.

Cao, Y., Grossberg, S., and Markowitz, J. (2011). How does the brain rapidly learn and reorganize view- and positionally-invariant object representations in inferior temporal cortex? *Neural Networks*, 24, 1050–1061.

Carpenter, G.A., and Grossberg, S. (1987). A massively parallel architecture for a self-organizing neural pattern recognition machine. *Computer Vision, Graphics, and Image Processing*, 37, 54-115.

Carpenter, G. A., and Grossberg, S. (1991). *Pattern recognition by self-organizing neural networks.* Cambridge, MA: MIT Press.

Chang, H.-C., Grossberg, S., and Cao, Y. (2014). Where's Waldo? How perceptual cognitive, and emotional brain processes cooperate during learning to categorize and find desired objects in a cluttered scene. *Frontiers in Integrative Neuroscience*, doi: 10.3389/fnint.2014.0043.

Dresp, B. (1997). On illusory contours and their functional significance. *Current Psychology of Cognition*, 16(4), 489–518.

Dresp, B., Durand, S., and Grossberg, S. (2002). Depth perception from pairs of overlapping cues in pictorial displays. *Spatial Vision*, 15, 255-276.

Dresp, B., Salvano-Pardieu, V., and Bonnet, C. (1996). Illusory form from inducers of opposite contrast polarity: Evidence for multistage integration. *Perception & Psychophysics,* 58, 111-124.

Dresp-Langley, B. (2014). On Galileo's visions: Piercing the spheres of the heavens by eye and mind. *Perception*, 43, 1280-1282. doi: 10.1068/p4311rvw.

Dresp-Langley, B. (2015). 2D geometry predicts perceived visual curvature in context-free viewing. *Computational Intelligence and Neuroscience*, Article ID: 708759, 1-9.

Fang, F., Boyaci, H., and Kersten, D. (2009) Border ownership selectivity in human early visual cortex and its modulation by attention. *The Journal of Neuroscience*, 29, 460-465.

Fang, L., and Grossberg, S. (2009). From stereogram to surface: How the brain sees the world in depth. *Spatial Vision*, 22, 45-82.

Fazl, A., Grossberg, S., and Mingolla, E. (2009). View-invariant object category learning, recognition, and search: How spatial and object attention are coordinated using surface-based attentional shrouds. *Cognitive Psychology*, 58, 1-48.

Foley, N.C., Grossberg, S., and Mingolla, E. (2012). Neural dynamics of object-based multifocal visual spatial attention and priming: Object cueing, useful-field-of-view, and crowding. *Cognitive Psychology*, 65, 77-117.

Gillam, B., Blackburn, S., and Nakayama, K. (1999). Stereopsis based on monocular gaps: Metrical encoding of depth and slant without matching contours. *Vision Research*, 39, 493-502.

Grossberg, S. (1984). Outline of a theory of brightness, color, and form perception. In E. Degreef & J. van Buggenhaut (Eds.), *Trends in Mathematical Psychology*, (pp. 59–85). Amsterdam: North-Holland.

Grossberg, S. (1994). 3D vision and figure-ground separation by visual cortex. *Perception and Psychophysics*, 55, 48-120.

Grossberg, S. (1997). Cortical dynamics of three-dimensional figure-ground perception of two dimensional figures. *Psychological Review*, 104, 618-658.

Grossberg, S. (1999). How does the cerebral cortex work? Learning, attention and grouping by the laminar circuits of visual cortex, *Spatial Vision,* 12, 163-186.

Grossberg, S. (2009). Cortical and subcortical predictive dynamics and learning during perception, cognition, emotion, and action. *Philosophical Transactions of the Royal Society of London*, 364, 1223-1234.

Grossberg, S. (2013). Adaptive Resonance Theory: How a brain learns to consciously attend, learn, and recognize a changing world. *Neural Networks,* 37, 1-47.

G      Grossberg, S. (2015). Cortical dynamics of figure-ground separation in response to 2D pictures and 3D scenes: How V2 combines border ownership, stereoscopic cues, and Gestalt grouping rules. *Frontiers in Psychology*.
http:       //dx.doi.org/10.3389/fpsyg.2015.02054.

Grossberg, S. , and Mingolla, E. (1985)

Grossberg, S., and Howe, P.D.L. (2003). A laminar cortical model of stereopsis and three-dimensional surface perception. *Vision Research,* 43, 801-829.

Grossberg, S., and McLoughlin, N.P. (1997). Cortical dynamics of 3-D surface perception:Binocular and half-occluded scenic images. *Neural Networks*, 10*,* 1583-1605.

Grossberg, S., and Pinna, B. (2012). Neural dynamics of Gestalt principles of perceptual organization: From grouping to shape and meaning. *Gestalt Theory*, 34, 399-482.

Grossberg, S., Srinivasan, K., and Yazdabakhsh, A. (2011). On the road to invariant object recognition: How cortical area V2 transforms absolute to relative disparity during 3D vision. *Neural Networks*, 24, 686-692.

Grossberg, S., Srinivasan, K., and Yazdanbakhsh, A. (2014). Binocular fusion and invariant category learning due to predictive remapping during scanning of a depthful scene with eye movements. *Frontiers in Psychology: Perception Science,* doi: 10.3389/fpsyg.2014.01457.

Grossberg, S., and Swaminathan, G. (2004). A laminar cortical model for 3D perception of slanted and curved surfaces and of 2D images: development, attention and bistability. *Vision Research*, 44, 1147-1187.

Grossberg, S., and Yazdanbakhsh, A. (2005). Laminar cortical dynamics of 3D surface perception: stratification, transparency, and neon color spreading. *Vision Research*, 45, 1725-1743.

Grossberg, S., Yazdanbakhsh, A., Cao, Y., and Swaminathan, G. (2008). How does binocular rivalry emerge from cortical mechanisms of 3-D vision? *Vision Research*, 48, 2232-2250.

He, Z. J., and Ooi, T. L. (1998) Illusory contour formation affected by luminance polarity. *Perception*, 27, 313-335.

Kapadia, M. K., Ito, M., Gilbert, C. D., & Westheimer,G. (1995). Improvement in visual sensitivity by changes in local context: Parallel studies in human observers and in V1 of alert monkeys. Neuron, 15, 843–856.

Kelly, F. J., and Grossberg, S. (2000). Neural dynamics of 3-D surface perception: Figure-ground separation and lightness perception. *Perception and Psychophysics,* 62,1596-1619.

Leveille, J., Versace, M., and Grossberg, S. (2010). Running as fast as it can: How spiking dynamics form object groupings in the laminar circuits of visual cortex. *Journal of Computational Neuroscience*, 28, 323-346.

Mathews, N., and Welch, L. (1997). The effect of inducer polarity and contrast on the perception of illusory figures. *Perception,* 26, 1431 – 1443.

McLoughlin, N.P. and Grossberg, S. (1998). Cortical computation of stereo disparity. *Vision Research*, 38, 91-99.

Nakayama, K., and Shimojo, S. (1990). da Vinci stereopsis: depth and subjective occluding contours from unpaired image points. *Vision Research*, 30, 1811-1825.

O'Herron, P., and von der Heydt, R. (2009). Short-term memory for figure-ground organization in the visual cortex. *Neuron,* 61 (5), 801-809.

O'Herron, P., and von der Heydt, R. (2011). Representation of object continuity in the visual cortex. *Journal of Vision,* 11, 12. doi: 10.1167/11.2.12.

Pinna, B., and Grossberg, S. (2006). Logic and phenomenology of incompleteness in illusory figures: New cases and hypotheses. *Psychofenia*, 9, 93-135.

Polat, U., & Norcia, A. M. (1996). Neurophysiological evidence for contrast dependent long-range facilitation and suppression in human visual cortex. *Vision Research*, 36, 2099–2109.

Prazdny, K. (1983). Illusory contours are not caused by simultaneous birghtness contrast. *Perception & Psychophysics*, 34, 403-404.

Prazdny, K. (1985). On the nature of inducing forms generating perceptions of illusory contours. *Perception & Psychophysics*, 37, 237-242.

Qiu, F. T., Sugihara, T., and von der Heydt, R. (2007). Figure-ground mechanisms provide structure for selective attention. *Nature Neuroscience* 10 (11), 1492-1499.

Qiu, F. T., and von der Heydt, R. (2005). Figure and ground in the visual cortex: V2 combines stereoscopic cues with Gestalt rules. *Neuron* 47(1), 155-166.

Rubin, E. (1921). *Visuell Wahrgenommene Figuren: Studien in psychologischer Analyse*. Kopenhagen: Gyldendalske.

Shapley, R., and Gordon, J. (1985). Non-linearity in the perception of form. *Perception & Psychophysics*, 37, 84-88.

Spehar, B. (2000). Degraded illusory contour formation with non-uniform inducers in Kanizsa configurations: the role of contrast polarity. *Vision Research*, 40, 2653-2659.

Spehar, B., and Clifford, C. W. G. (2003). When does illusory contour formation depend on contrast polarity? *Vision Research*, 43, 1915-1919.

Spillmann, L., Dresp-Langley, B. and Tseng, C. (2015) Beyond the classical receptive field: The effect of contextual stimuli. *Journal of Vision*, **15**, 1–22.

Tse, P. U. (2005). Voluntary attention modulates the brightness of overlapping transparent surfaces. *Vision Research*, 45, 1095-1098.

Tzvetanov, T., & Dresp, B. (2002). Short- and long-range effects in line contrast detection. *Vision Research, 42*, 2493-2498.

Wehrhahn C., and Dresp B. (1998). Detection facilitation by collinear stimuli in humans: Dependence on strength and sign of contrast. *Vision Research, 38*, 423-428.

Yazdanbakhsh, A., and Grossberg, S. (2004). Fast synchronization of perceptual grouping in laminar visual cortical circuits. *Neural Networks*, 17, 707-718.

Su, Y.R., He, Z. J., and Ooi, T. L. (2010) Boundary contour-based surface integration affected by color. *Vision Research*, 50, 1833-1844.

Zhang, N. R., and von der Heydt, R. (2010). Analysis of the context integration mechanisms underlying figure-ground organization in the visual cortex. *The Journal of Neuroscience* 30(19), 6482-6496.

Zhou, H., Friedman, H. S., and von der Heydt, R. (2000). Coding of border ownership in monkey visual cortex. *The Journal of Neuroscience* 20(17), 6594-6611.

# Figure Captions

**Figure 1 -** Two faces or a vase? In these variations on the famous reversible figures of Rubin (1921), with surface contrasts of opposite signs, the perceptual assignment of border ownership to foreground and background may be influenced by both shifts in spatial attention and prior learning of object categories.

**Figure 2** – Figure 2 here is organized in three parts: part 1) four Kanizsa configurations are shown in two columns on the top left. The two Kanizsa squares in the first column represent stimuli used in experiments on sign-invariant boundary detection by Shapley and Gordon (1985) and obey their criteria of sign-invariant boundary induction. The two Kanizsa squares in the second column show cases where single inducing elements are given locally opposing contrast signs, used as stimuli in experiments by Spehar (2000) and Spehar and Clifford (2003). These two configurations do not obey Shapley and Gordon's criteria of sign-invariant boundary induction. The strength of the illusory boundaries therein was reported to be less discriminable, and even more so when exposure duration was limited to less than 320 milliseconds (Spehar, 2000; Spehar and Clifford, 2003). An explanation of this finding in terms of FAÇADE properties is suggested on page 12, lines 372-375 here in our manuscript. Part 2) six Ehrenstein configurations with centrally induced surfaces are shown in the two columns on the top right here. The illusory surface in the centre was reported less perceptible when the radial inducing lines are fragmented (as in the Ehrenstein configurations here in the right column) and given locally opposing contrast signs (Dresp, Salvano-Pardieu and Bonnet, 1996; Spehar and Clifford, 2003). When all fragments share the same contrast sign (as in the configuration at the bottom of the left column), the famous 'O' illusion discovered by He and Ooi (1998) is perceived, which also exists in color (Yong, He and Ooi, 2010). This percept is abolished when the local contrast signs are of the opposite polarity (as in the configuration at the bottom of the right column). Part 3) The six visual configurations presented in the psychophysical experiment of this study here are shown at the bottom here. These six spatially discontinuous shape configurations were created using the criteria of Shapley and Gordon (1985) for sign invariant boundary completion and surface filling-in. They generate unambiguous figure-ground percepts of continuous surfaces in depth. In the upper row of these images, the outward-directed contrast edges make the central surface more likely to be seen as lying "behind" the surrounding surface, whereas in the lower row of images, the inward-directed edges make the central surface more likely to be seen as standing out "in front" of' the surround, as predicted by generic assumptions of FAÇADE and 3D Laminart and confirmed by the experimental data here.

**Figure 3 -** Probabilities of perceptual decisions for figure ("in front") or ground ("behind") assignment of the surface in the center of the images with fragmented edge contours, plotted as a function of the direction of the local edge contrasts and their contrast sign.

**Figure 4 –** The FACADE model macrocircuit. The illuminant-discounted inputs from the Right and Left Monocular Preprocessing stage, which is composed of center-surround cells, output to the Left and Right Monocular boundaries composed of simple cells via pathways 1. Left and Right Monocular Boundaries are binocularly fused via pathways 3. Pathways 4 and 5 complete these boundaries using bipole grouping at the Binocular Boundaries stage. Depthful binocular boundaries mutually interact with the Monocular Surfaces stage (pathways 6), where the closed boundaries are filled-in by the illuminant-discounted surface input. The attached boundaries to the successfully filled-in surfaces generate surface contour outputs signals. These signals strengthen the boundaries that induced them, and prune the redundant boundaries at the same positions and further depths (pathways 7). The Binocular Surfaces stage binocularly fuse excitatory inputs from the Left and Right Monocular Preprocessing stages (pathways 8) while surface pruning occurs of redundant feature contours at further depths (pathways 9). Boundary enrichment of the Binocular Boundaries occurs at the Binocular Surfaces and regulates surface filling-in there. Boundaries are enriched by adding boundaries at same positions from near depths to far depths. Due to surface pruning, the illuminant-discounted surface inputs that are contained by the enriched boundaries are pruned from the further depths where boundaries are added.

**Figure 5 -** 3D LAMINART model circuit diagram. This laminar visual cortical model consists of a boundary stream that includes V1 interblobs, V2 pale stripes (also called interstripes), and part of V4, and computes 3D perceptual groupings in different scales; and a surface stream that includes V1 blobs, V2 thin stripes, and part of V4, and computes 3D surfaces that are infused with lightness in depth. Both the boundary and surface streams receive illuminant-discounted signals from LGN cells with center-surround receptive fields, and both converge in V4, where visible 3D surfaces are consciously seen that are separated from their backgrounds. Models V2 and V4 also output to inferotemporal cortex (not shown), where object recognition takes place. Model V1 interblobs contain both monocular and binocular cells. Binocular simple cells become disparity-sensitive by binocularly matching

left and right scenic contours with the same contrast polarity in layer 3B before pooling opposite polarity responses at complex cells in layer 2/3A. Monocular and binocular boundary cells control filling-in of monocular 3D surfaces within V1 blobs. Closed boundaries can contain the filling-in process, and can send feedback to V1 interblobs that selectively strengthens the closed boundary components. Monocular and binocular V1 boundaries are pooled in V2. V2 pale stripes can complete 3D perceptual groupings while inhibiting false binocular matches using the disparity filter to solve the correspondence problem. These completed boundaries form compartments in the V2 thin stripes within which filling-in of monocular 3D surfaces occurs. Closed boundaries can contain the filling-in process and send surface-to-boundary surrface contour feedback signals to enhance their generative boundaries, while also suppressing redundant boundaries at the same positions and frrther depths. These conmpleted boundaries and filled-in surfaces complete the representations of partially occluded objects. They do not generate visible percepts, but can be recognized by activating inferotemporal cortex. Visible surfaces in which figures are separated in depth from their backgrounds are formed in V4. Here, left and right eye feature contour signals from the LGN are binocularly matched, while redundant feature contour signals are pruned at further depths by inhibitory signals from the thin stripes. Then the pruned feature contour signals induce filling-in of a visible surface percept within enriched binocular boundaries. V4 emits output signals that lead to recognition and grasping of unoccluded parts of opaque surfaces - Reproduced with permission from Fang and Grossberg (2009).

**Figure 6 -** (a) T-Junction Sensitivity: (left panel) and T-junction in an image (middle panel). Bipole cells provide long-range cooperation (+), and work together with inhibitory interneurons that provide cells provide short-range competition (-) - (right panel). An end gap in the vertical boundary arises because, for cells near where the top and stem of the T come together, the top of the T activates bipole cells along the top of the T more than bipole cells are activated along the T stem. As a result the stem boundary gets inhibited whereas the top boundary does not - Reprinted with permission from Grossberg (1997) - (b) Necker cube. This 2D picture can be perceived as either of two 3D parallelograms whose shapes flip bistably through time. (c) When attention switches from one circle to another, that circle pops forward as a figure and its brightness changes. See Grossberg and Yazdanbakhsh (2005) for an explanation - Reprinted with permission from Tse (2005).

**Figure 7** - The top row illustrates how, at a prescribed depth, a closed boundary contour abuts an illuminant-discounted feature contour. When this happens, the feature contours can fill-in within the closed boundary. The bottom row (left panel) depicts how filling-in of the feature contours is contained by this closed boundary contour, thereby generating large contrasts in filled-in activity at positions along the boundary contour. Contrast-sensitive surface contour output signals can then be generated in response to these large contrasts. The bottom row (right panel) depicts a boundary contour that has a big hole in it at a different depth. Feature contours can spread through such a hole until the filled-in activities on both sides of the boundary equalize, thereby preventing contrast-sensitive surface contour output signals from forming at such boundary positions - Reprinted with permission from Grossberg (2015).

**Figure 8** - A closed boundary can form at Depth 1 by combining a binocular vertical boundary at the left side of the square with three monocular boundaries that are projected along the line of sight to all depths. Surface contour output signals can thus be generated by the FIDO at Depth 1, but not the FIDO at Depth 2. The Depth 1 surface contours excite, and thereby strengthen, the boundaries at Depth 1 that controlled filling-in at Depth 1. These surface contours also inhibit the redundant boundaries at Depth 2 at the same positions. As a result, the pruned boundaries across all depths, after the surface contour feedback acts, can project to object recognition networks in inferotemporal cortex to facilitate amodal recognition, without being contaminated by spurious boundaries - Reprinted with permission from Grossberg (2015).

**Figure 9** - Initial steps in generating a 3D percept of figures at different depths in response to a 2D picture with particular occlusion. (a) This figure is composed of three abutting rectangles but generates a percept of a vertical rectangle that partially occludes a horizontal rectangle. Due to mechanisms described in the text, the boundary of the vertical rectangle is separated onto a near depth D1 and achieves border ownership of its shared boundaries with the two smaller rectangles. The remaining boundaries are separated onto a slightly further depth D2, where they can use bipole completion to complete the boundary of the partially occluded horizontal rectangle (dotted lines). This picture does not show the boundary fragments at depth D1 in which end gaps have been generated. The text and Figure 10 propose how end gap boundaries are eliminated. (b) This figure is composed of two abutting rectangles. Although there is no completion of the horizontal rectangle behind the vertical

rectangle, a 3D percept can nonetheless be generated using the same mechanisms - Adapted with permission from Grossberg (1997).

**Figure 10 -** How spurious end gap boundaries are eliminated. This figure illustrates how spurious end gap boundaries are eliminated from the near depth D1 in the 3D percept that is generated by the 2D picture in Figure 9b. In this case, the end gap boundaries at depth D1 in (a) cannot be eliminated, as they can in response to the percept generated by Figure 9a, by the disparity filter in V2 after surface contour feedback strengthens closed boundaries at the pale stripes from thin stripes. This is true because the boundary at depth D2 is not closed; see (a). On the other hand, this boundary is closed by boundary enrichment in V4; see (b). As a result, top-down attention from the filled-in surfaces in V4 (see (c)) can strengthen the boundaries of closed regions in V2 (see thicker lines in (d)). After this happens, the disparity filter in V2 can eliminate the end gap boundary at depth D1.

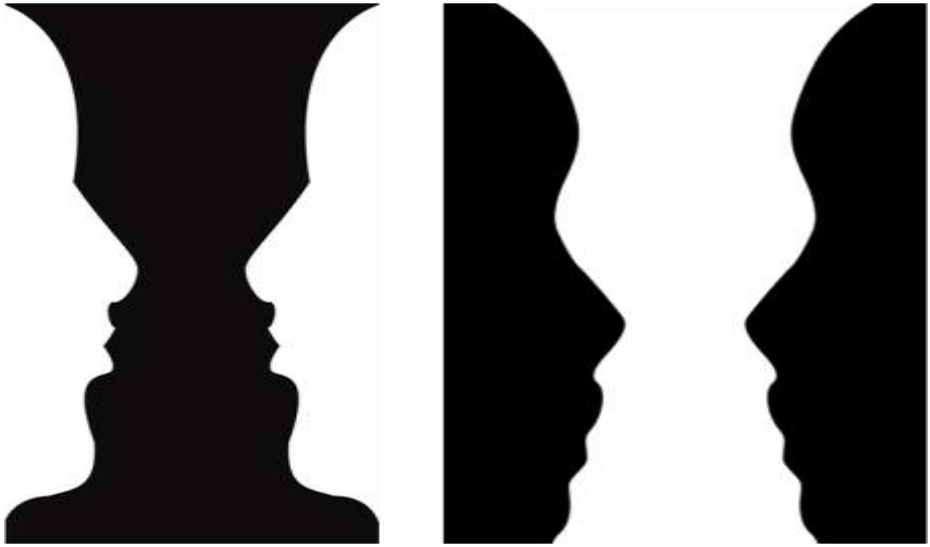

Figure 1

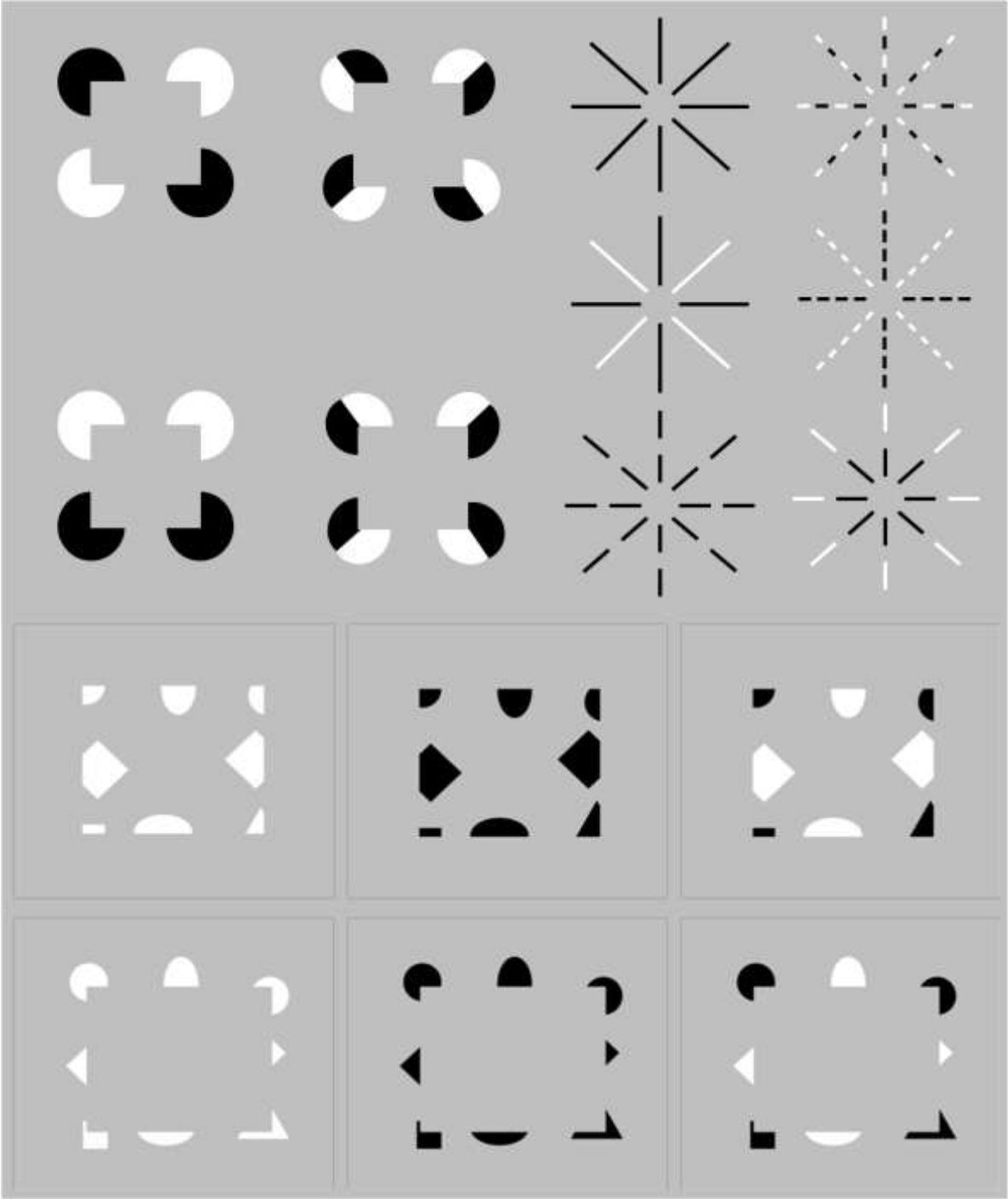

Figure 2

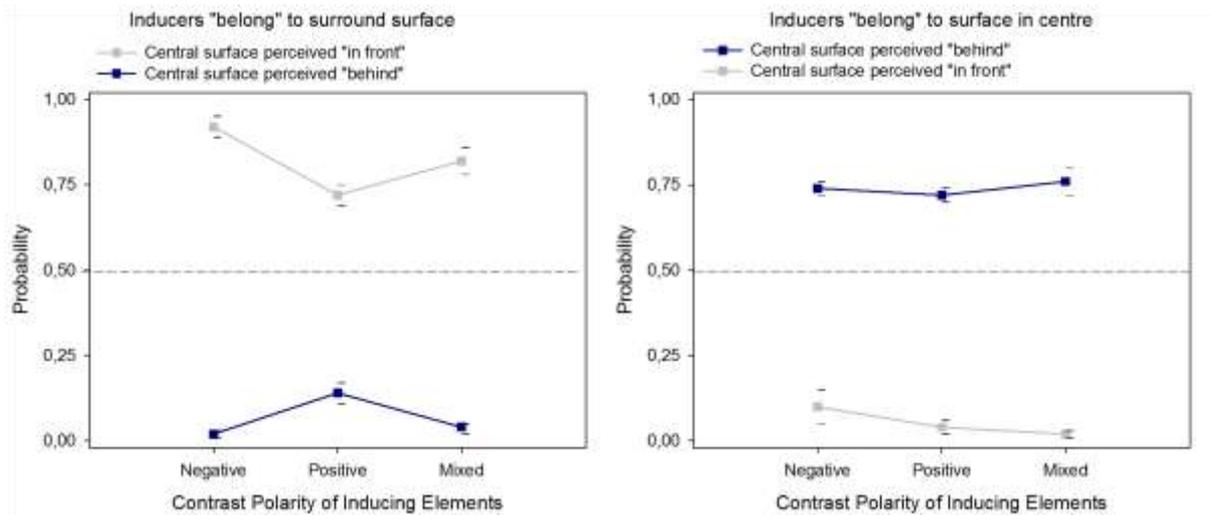

Figure 3

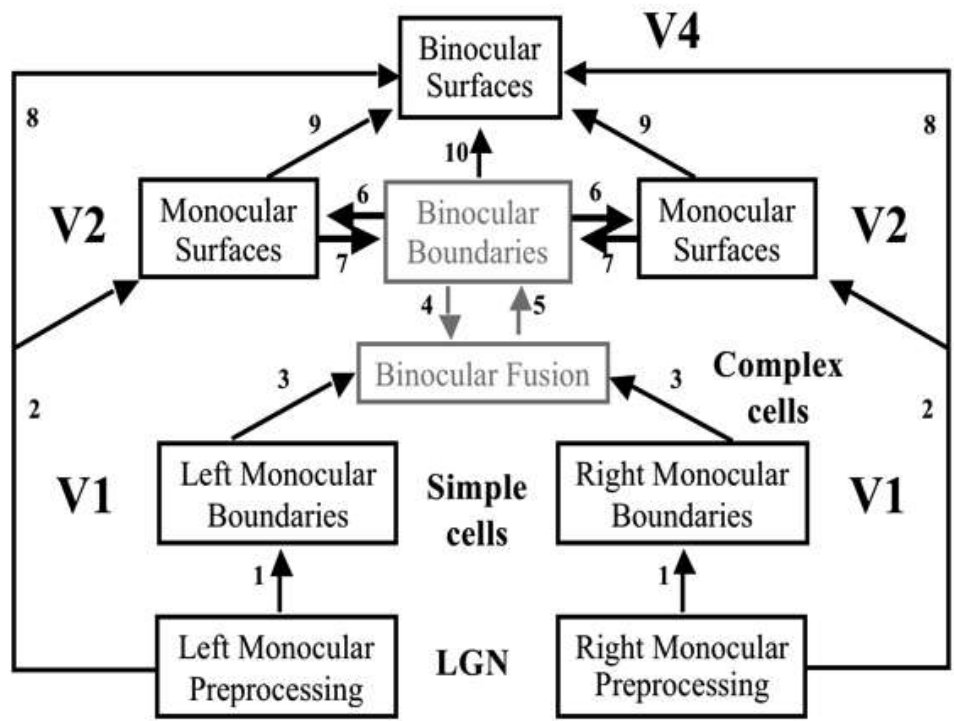

Figure 4

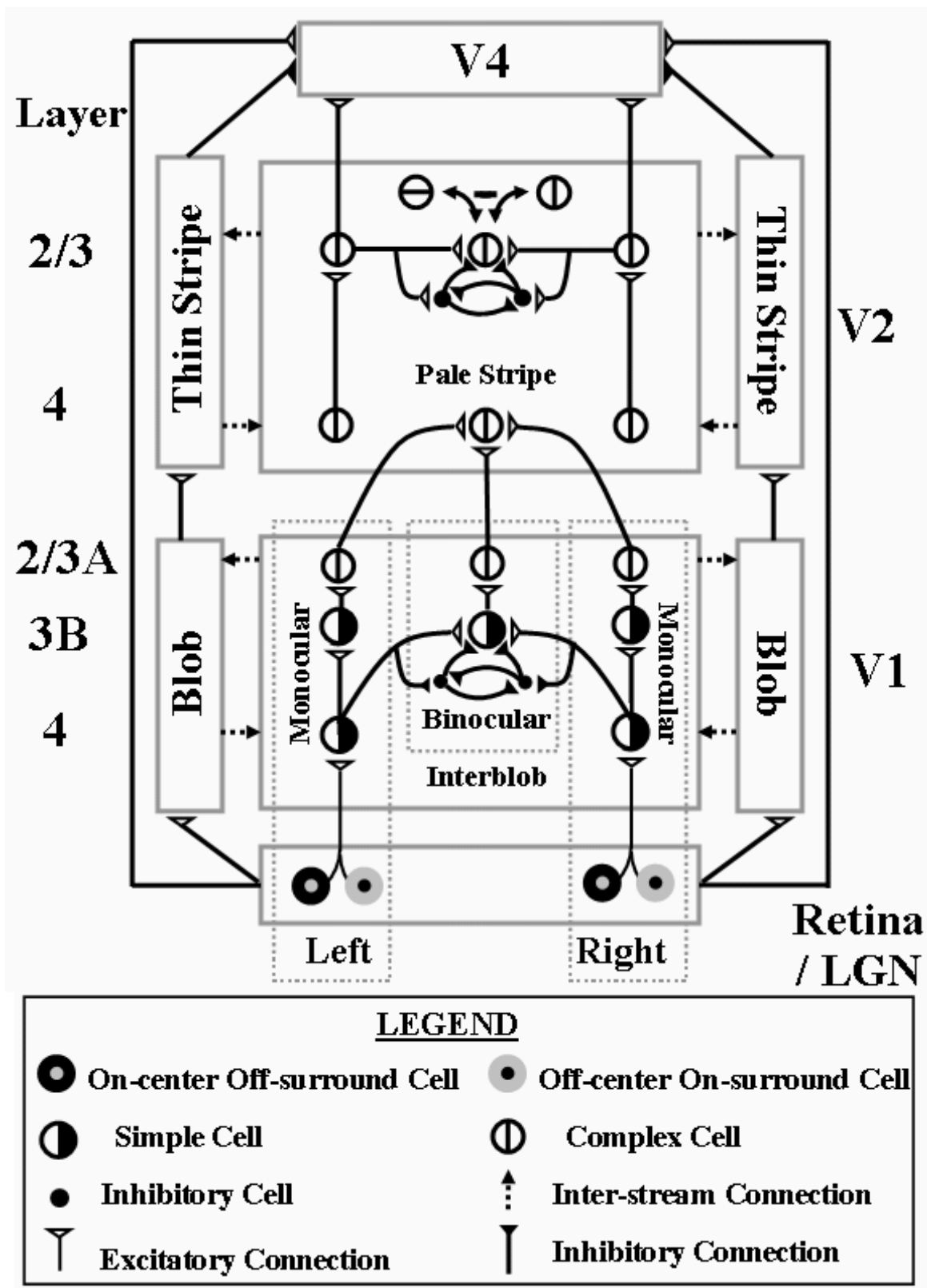

Figure 5

**T-JUNCTION SENSITIVITY**

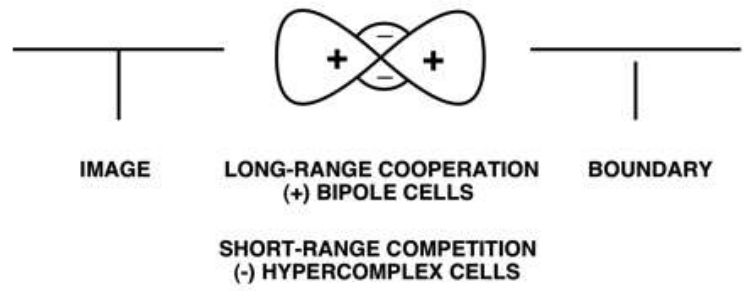

IMAGE     LONG-RANGE COOPERATION     BOUNDARY
(+) BIPOLE CELLS

SHORT-RANGE COMPETITION
(-) HYPERCOMPLEX CELLS

(a)

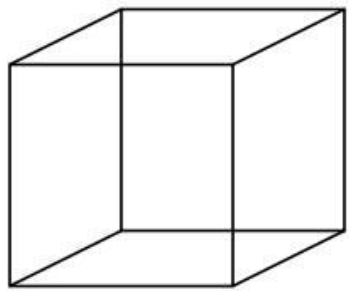

(b)

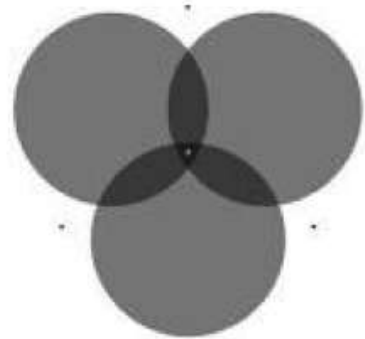

(c)

Figure 6

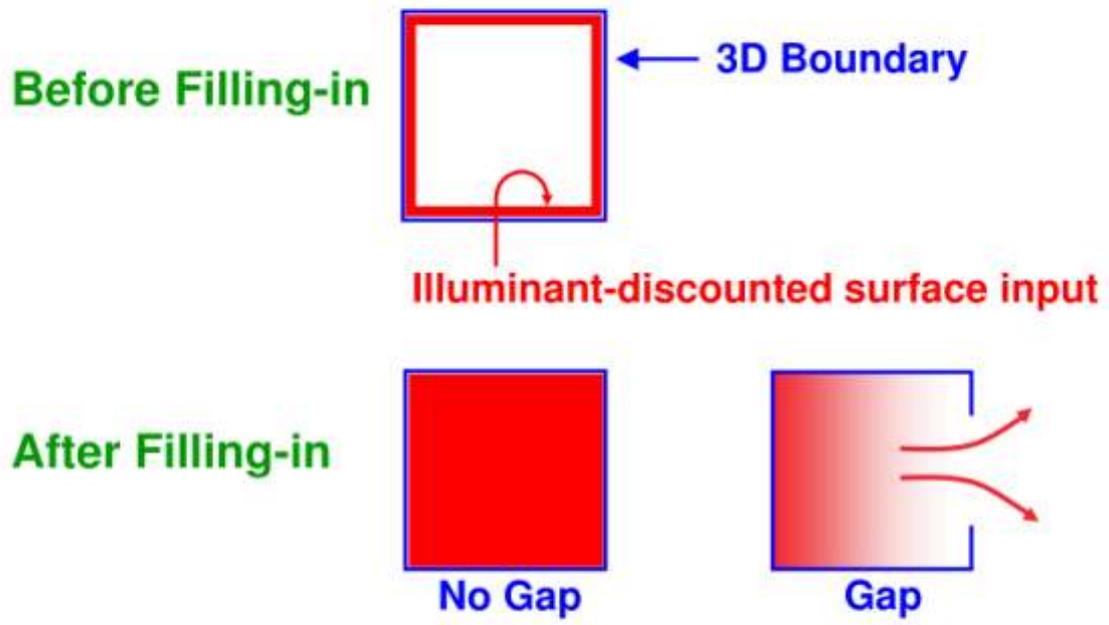

Figure 7

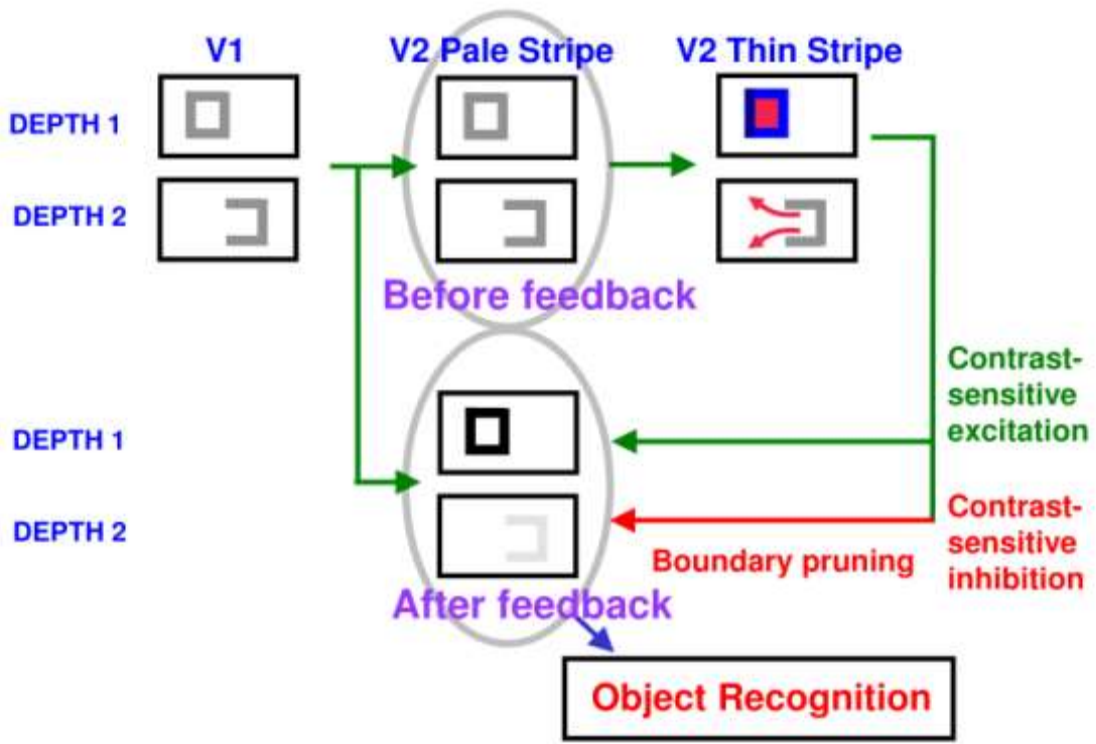

Figure 8

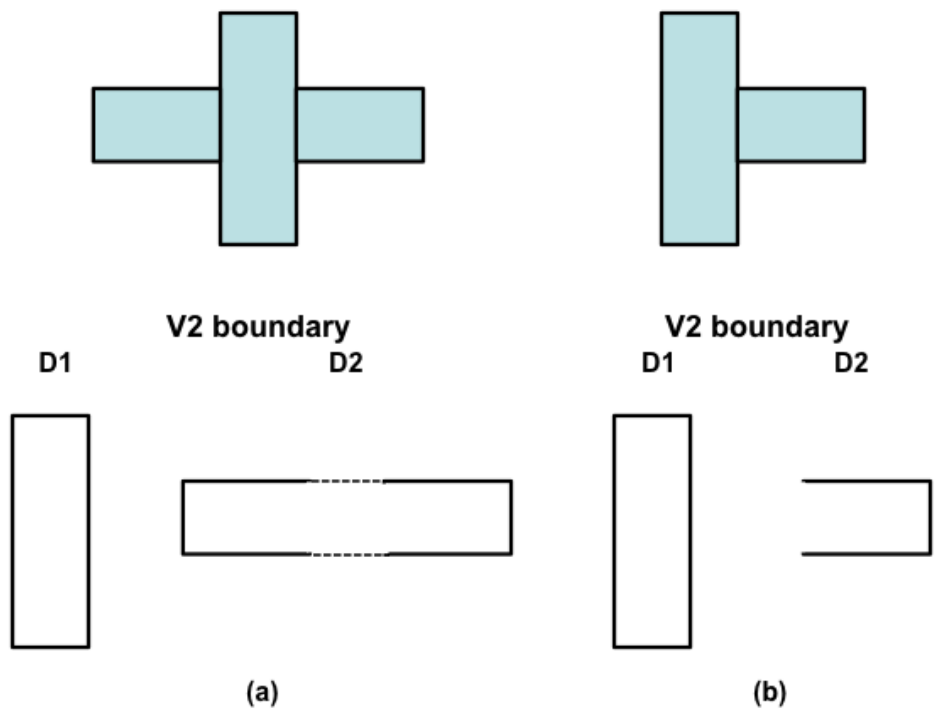

Figure 9

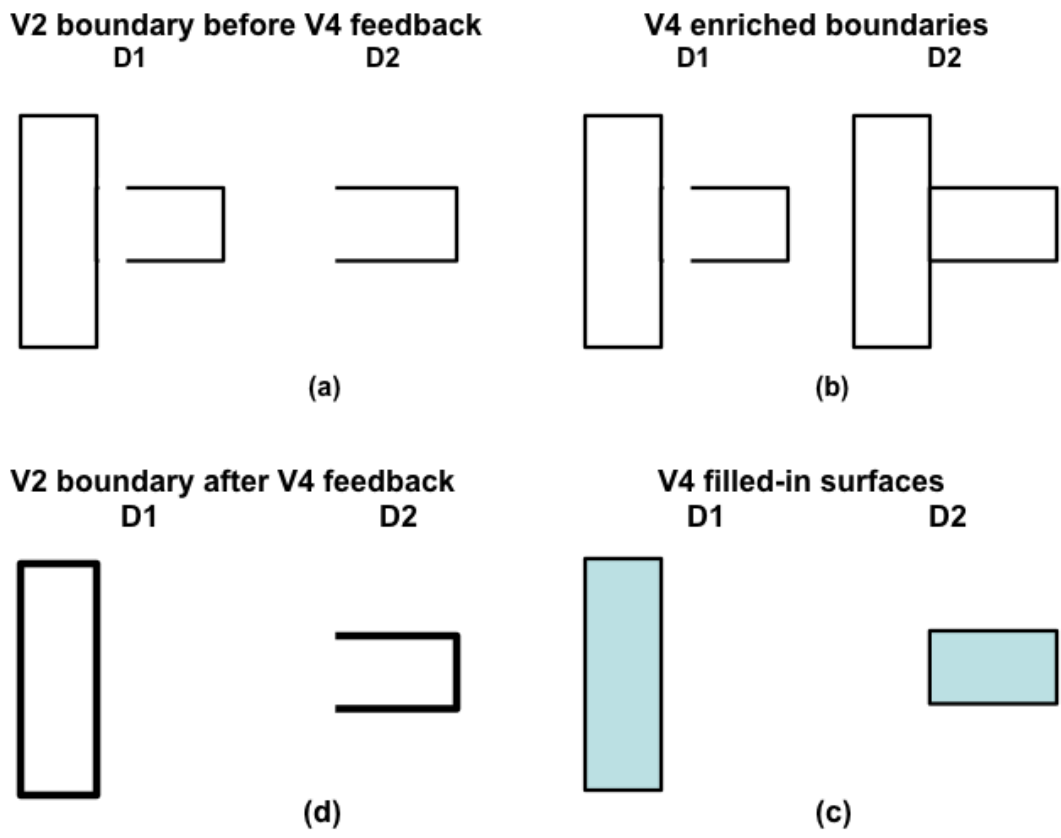

Figure 10